# Inventions on GUI for Eye Cursor Control Systems
## A TRIZ based analysis

**Umakant Mishra**

Bangalore, India

http://umakantm.blogspot.in

**Contents**

## 1. Introduction

### 1.1 Need for an Eye Cursor Control system

Computer interfaces such as mouse, track balls, and light pens etc. provide users with a way of controlling and manipulating the display of information on a computer screen. Although a mouse is popular and effective it has disadvantages typically in the circumstances where;

- ⇨ The user wants to use hands for other purpose, e.g., the user wants to use keyboard for typing simultaneously.

- ⇨ The user requires a hands-free environment

- ⇨ The user is physically challenged or handicapped as to the use of his or her hands.

For example, a medical doctor who is performing medical or surgical procedure with his hands may benefit from such an eye ball controlled system to scroll and view the CAT Scan, S-Ray images, Patient Charts or reference medical text books etc.

Similarly automobile mechanics or electrical technician or assembly line workers who are working on their primary job with their hands can simultaneously refer to a manual or other special instructions without moving their hands out.

### 1.2 Eye Tracking Devices

Eye tracking devices are based on Video or Image Processing. There may be two different types of Eye Tracking devices, one with an attachment to human head, another without any attachment.

The tracking device with an attachment requires the device to be attached to human head in order to track the movement of the user's eyes. But users typically don't prefer to wear attachments on them in order to track the movement of their eyes. This could be because the attachments are uncomfortable to wear or not aesthetically pleasing or fashionable.

The non-attached eye tracking systems place a device/ camera on the monitor. An example of such a non-attached device is developed by LC Technologies (http://www.lctinc.com/doc/ecs.htm). However, a problem with these non-attached eye-tracking devices is that they allow the movement of the user's head only within a limited range.

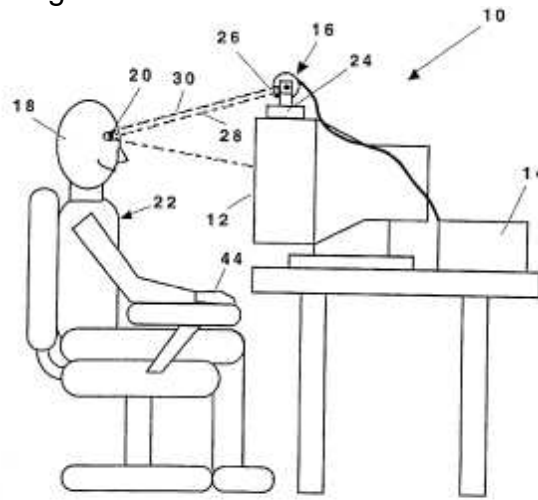

Picture from US Patent 6351273

A better eye-tracking device should allow head movement without head attachment. US Patent 6351273 (illustrated below) provides a non-attached eye tracking system, which works without restricting the movement of user's head.

### 1.3 Eye Cursor Control systems

The eye mouse devices use movement of the eyes of the user to accomplish the movement of the cursor. Some earlier Eye Mouse Devices still require Manual or Foot Activation to control the display of information, text, images or data. The main disadvantage is that they still require the use of user's hand or foot.

The Eye Mouse Devices were further improved to control the cursor movement by the movement of the user's eyes without needing any support from hand or foot. These devices allow all kinds of operations including movement of cursor, activation of menus and selection of text etc. (US Patent 4836670, 4950069, 4973149, 5345281) by the user dwelling, gazing or staring at a desired activation region for a predetermined amount of time.

However, these systems are typically slow as the user as to wait atleast for a predefined period before the selection, scrolling or specific operation is activated. Recent inventions try to overcome these limitations by providing alternative mechanism, e.g., US Patent 6351273 illustrated below provides a device for automatic scrolling.

### 1.4 Problems in eyeball tracking systems

There are problems in the known eyeball tracking mechanisms. Such as:

⇒ Using a head-mount is not comfortable, but without a head-mount it is difficult to track eyeball as the user might move his head.

⇒ The eyeball tracking mechanisms are not accurate. The accuracy is even worse in small display screens or limited screens containing too many GUI elements.

⇒ The user may have physical limitations to focus the eyeball exactly on the GUI element. The mechanism may not work well if the user focuses near the edge of the GUI element.

# 2. Inventions on GUI for systems based on eyeball tracking

### 2.1 Assisting selection of GUI elements (6323884)

**Background problem**

Eyeball tracking instruments track the eye movement of a user and select the elements of the GUI depending on he eye ball position. There are some problems in this mechanism, especially the accuracy of measuring eyeball movement. Besides the mechanism does not work well when the user focuses near an edge of the GUI element. There is a need to assist and validate the user selection process.

**Solution provided by the invention**

Bird et al. found this solution patented Nov 2001 (Patent 6323884, assigned by IBM). According to the invention, the software identifies the potential GUI elements that the user may interact with. A predefined set of characteristics for the identified GUI elements are compared to predict which of the GUI elements are likely to be selected by the user. The result of this prediction is indicated by moving the pointer to the predicted GUI element.

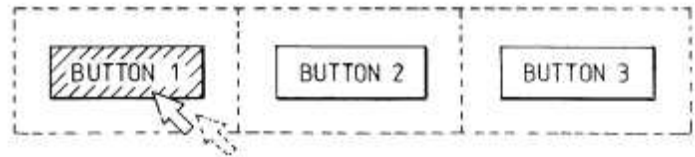

In case the pointer lies in the border, the invention pulls it on to the nearest selectable item. This action makes the user know about the effects of his selection.

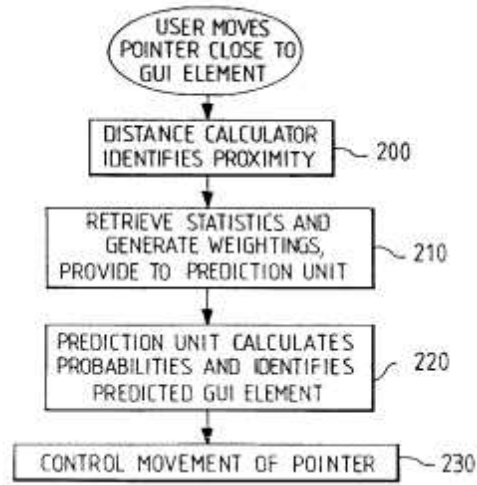

**TRIZ based analysis**

If the pointer lies in the border, the invention pulls it on to the nearest selectable item (Principle-8: Counterweight, Principle-16: Partial or excessive action).

This action gives a feedback to the user about his selection and allows him to rectify if not correct. (Principle-23: Feedback)

Besides the system prompts possible selectable items based on previous activities of the user (Principle-10: Prior Action).

## 2.2 System and methods for controlling automatic scrolling of information on a display or screen (6351273)

### Background problem

Eye tracking systems are very useful in a hands-free environment where the user does not have hands to operate a mouse or wants to use the hands for other activities like using a keyboard. Although there have been several inventions in this field, each has its own limitations.

This situation creates is a need for a hands-free eye-controlled scrolling device for computer systems. There is also a need to provide an automatic scroll control device for automatically scrolling the display of information, text, data, images etc. on a computer display screen to provide a hands-free environment.

### Solution provided by the invention

Patent 6351273 (invented by Lemelson et al., issued Feb 2002) discloses a method of automatic scrolling of information on a computer display by tracking the position of user's head and eye using a computer-gimbaled sensor.

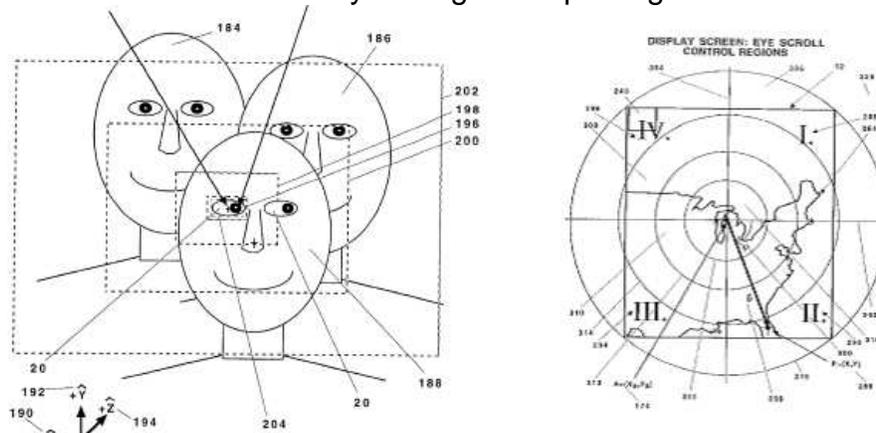

The gimbaled sensor system tracks the eye of the user and an eye gaze direction determining system determines the eye gaze direction. An automatic scrolling system scrolls the screen based on calculated screen gaze coordinates of the eye of the user.

### TRIZ based analysis

The invention uses an eye tracking mechanism to scroll the screen instead of conventional mouse like pointers (Principle-28: Mechanics substitution).

As the head-mounts are uncomfortable to the user, the invention uses a non-attached mechanism, i.e., no attachments are mounted on user's head (Principle-2: Taking out).

The method discloses an automatic scroll control mechanism (Principle-25: Self service).

The method segments the screen control region into a number of concentric circles (Principle-1: Segmentation, Principle-14: Curve).

The invention discloses an algorithm of converting the user's eyeball movement to find screen gaze coordinates (Principle-36: Conversion).

## 3. Summary

Operating a GUI through eyeball is a complex mechanism and not used as often as mouse or trackball. But there are situations where eye-mouse devices can play a tremendous role especially where the hands of the user are not available or busy to perform other activities. The difficulties of implementing an eye-cursor control system are many. The article illustrates some inventions on eye-cursor control system, which attempt to eliminate the difficulties of the prior art mechanisms.